# A Directional Monitoring Approach of Sequential Incomplete Wind Power Curves with Copula-based Variational Inference

Peng Wang[1], Yanting Li[1,*], Fugee Tsung[2,3]

[1] School of Mechanical Engineering, Shanghai Jiao Tong University, Shanghai, China

[2] Department of Industrial Engineering and Decision Analytics, the Hong Kong University of Science and Technology, Kowloon, Hong Kong

[3] Information Hub, Hong Kong University of Science and Technology (Guangzhou), Guangzhou, China

**Abstract:** Wind turbines often work under complex conditions which result in performance degradation. Accurate performance degradation monitoring is essential to ensure the reliable operation of wind turbines and reduce the maintenance costs. Wind turbine power curve monitoring is an effective way to detect performance degradation. However, due to the intermittency and fluctuation of wind speed, the wind speed range varies at different time periods, making power curves difficult to compare. Motivated by this, we proposed copula-based variational inference framework and used it to establish a sequential incomplete wind power curve estimation algorithm. First, a monotone power curve is constructed based on copula-based variational inference and integrated spline regression model. Besides, the prior distribution of model parameters are sequentially updated. Then, a directional control chart based on a new statistic named KL-divergence factor is constructed to monitor wind turbine performance degradation. The real data of a wind farm in the east of the United Kingdom shows that the proposed method can both improve the accuracy of wind turbine power curve modeling and monitor wind turbine performance degradation more precisely and comprehensively than the existing approaches.

# 1. Introduction

Wind energy is a renewable, pollution-free, and widely distributed energy source that has received increasing attention. Wind power plants generate electricity by equipping a series of wind turbines based on wind conditions, surrounding terrain, transmission routes, and other site selection considerations. The electricity generated by each turbine is transmitted to the substation and then to the power grid. The cumulative installed capacity of wind turbines continues to grow, increasing from 198GW in 2010 to 837GW in 2021 [1].

However, due to fluctuations in wind and energy demand, wind turbines are subject to complex conditions under alternating loads. The harsh working conditions are more likely to result in performance degradation, reduction of wind power output and further increase of the operation and maintenance cost. Accurate performance degradation monitoring is essential to ensure the reliable operation of wind turbines. Power curve monitoring focuses on detecting the changes in the relationship between power output and explanatory variables. Anomalies may occur if the functional relationship changes.

Power curve monitoring is a typical profile monitoring problem. Profiles can be categorized as parametric model and non-parametric model. Parametric model assumes that response variable is correlated with explanatory variables, and regression models are often used to explain this relationship [2, 3, 4, 5], such as univariate linear regression, multivariate regression, polynomial regression, generalized linear regression, nonlinear regression, etc. Detecting changes in the regression coefficients is the main task.

However, fitting parametric models of the in-control and out-of-control profile data is not always successful. Thus, some researchers proposed non-parametric profile models, including local kernel regression (LKR) [6, 7], functional data analysis (FDA) [8, 9], functional principal component analysis (FPCA) [10], gaussian process regression (GPR) [11], etc. Qiu et al. [6] introduced local linear kernel smoothing into an exponentially weighted moving average (EWMA) control chart, and used a non-parametric mixed-effect model to describe in-profile correlations. Zou et al. [7] combined multivariate exponential weighted moving average control chart (MEWMA) and generalized likelihood ratio test (GLRT) based on non-parametric regression to carry out online monitoring of changes in regression relations and profile variance. Functional data analysis including wavelet transform, B-spline transform, etc, regards profile data as a continuous function, and converts it into a certain feature space to obtains feature coefficients. Zhou [8] combined statistical process control with Haar wavelet transformation to not only detect process changes but estimate the amplitude of mean shift. Chang et al. [9] used discrete wavelet transform to separate the variance or noise from the profile, B-splines to define the shape of the profile, and combined with the Hoteling T2 control chart to monitor the mean shift or shape change. Colosimo et al. [10] investigated the application of principal component analysis in profile data analysis and explored which types of profile features were easy to obtain interpretable PCs. Due to its flexibility and conciseness, GPR shows good modeling performance. For example, Zhang et al. [11] used GPR to characterize correlations within profiles, which focused on Phase II monitoring of linear trends and inner profile correlations and established two multivariate Shewhart control charts.

Referring the profile monitoring methods, existing power curve monitoring approaches compare the latest power curve with the reference power curve under normal condition either by transforming power curve to a scalar metric[12, 13], or comparing deviations between two power curves [14], or monitoring model parameters [Long, Wang 2015]. AEP [12] and the power coefficient [13] are common scalar metrics that measure the properties of power curves. If the AEP or power coefficient of one wind turbine is inferior to that of the reference power curve built under the normal working condition, then this wind turbine may suffer from performance degradation. Ding et al. [14] discussed wind turbine performance monitoring based on the power curves in both the spatial and temporal domain. Long et al. [15] applied the Hotelling $T^2$ and generalized variance charts to monitor power curve parameters. Kusiak et al. [16] investigated three reference curves including the power curve, rotor curve and blade pitch curve, and applied the Hotelling T2 chart to monitor the skewness and kurtosis of these curves.

However, due to the intermittency and fluctuation of wind speed, the wind speed range varies at different time periods (Fig. 1), which leads to significant differences in the shape of the power curves.

The basic idea of the new sequential incomplete power curve directional monitoring approach consists of three main parts. Firstly, a new variational inference named Copula-based Variational Inference (CVI) is proposed to obtain more accurate posterior distribution of latent variables without severely underestimating posterior variance as Mean Field Variational Inference (MFVI). Secondly, a monotone power curve based on CVI, and Integrated Spline (I-Spline) is proposed to describe the wind power generation efficiency with incomplete samples for sequential data segment updating. Finally, an online power curve monitoring method for directional change of mean vector is proposed to detect performance degradation of wind turbines.

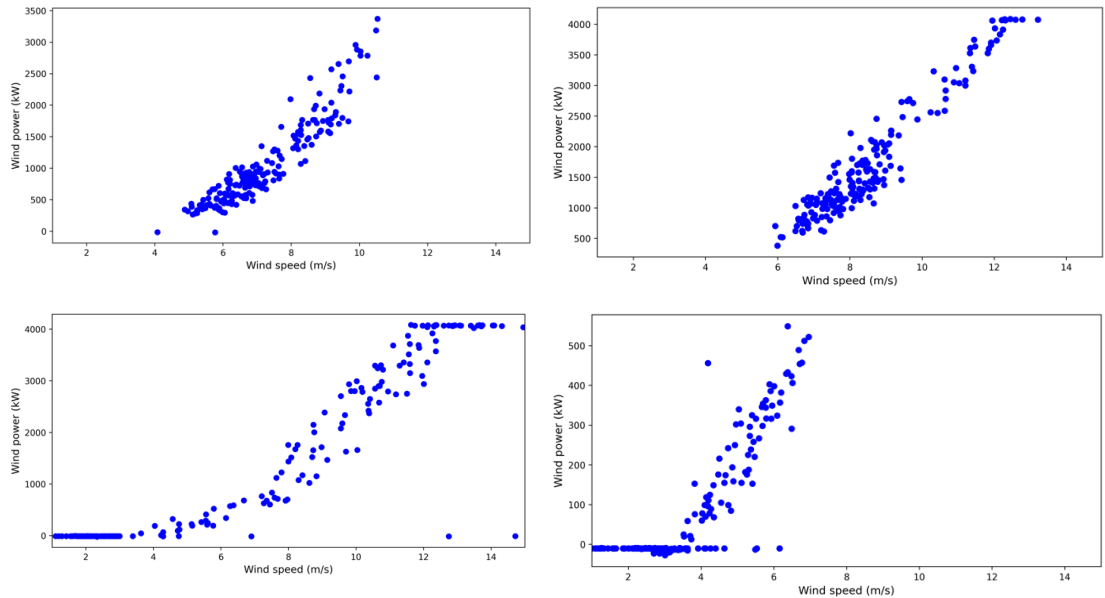

Fig 1. Power curves for the same wind turbine in different time periods

The proposed method enjoys the following innovation:

1) Unlike traditional profile construction methods that entails a large amount of data samples, the power curve of wind turbines can be sequentially updated with limited data samples accurately based on CVI. This solves the inaccuracy problem of Mean Field Variational Inference (MFVI) and can improve the response speed of performance degradation detection.

2) The power curve monitoring approach proposed is online and directional. It has lower computational burden and is more sensitive to directional changes in power curve parameters.

The rest of the paper is organized as follows. Section 2 described the directional monitoring approach of sequential incomplete power curves based on CVI in detail. In Section 3, the performance analysis in a real-world dataset is presented. We finally conclude in Section 4.

## 2. Preliminaries

### 2.1 Copulas

To describe the dependence among random variables, Sklar [17] proposed the copula theory. A copula is a multivariate distribution function whose marginal distribution follows a uniform distribution $\mathbb{U}(0,1)$. Let $\Theta = [\Theta_1, \dots, \Theta_p]'$ be the $p$-dimensional continuous random variable that we want to model their joint probability density function (PDF) $f(\theta_1, \dots, \theta_p)$ and let $F_1(\theta_1), \dots, F_p(\theta_p)$ be the cumulative distribution functions (CDF) of these random variables respectively, according to Sklar's theorem, $f(\theta_1, \dots, \theta_p)$ can be decomposed into $p$ univariate marginal distributions and a copula function. The joint distribution of random variables can be modeled by estimating marginal distributions independently of the copula [18], which makes it popular in high-dimensional data analysis. Mathematically, the joint cumulative distribution function (CDF) for p-dimensional random vector $X$ can be written as,

$$F(\theta_1, \dots, \theta_p) = C\left(F_1(\theta_1), \dots, F_p(\theta_p)\right) \tag{1}$$

where $C(\cdot)$ denotes a copula function; $F_i(\theta_i)$ ($i = 1, \dots, p$) represents the $i$-th marginal CDF satisfying

$$F_i(\theta_i) = \int_{-\infty}^{\theta_i} f_i(t) dt \tag{2}$$

where $f_i(\theta_i)$ is the PDF of the $i$-th variable. If $F_i(\theta_i), i = 1, \dots, p$ are continuous, then copula $C$ is unique. If $C$ is differentiable, the copula density $c$ can be derived by the following differentiation,

$$c\left(F_1(\theta_1), \dots, F_p(\theta_p)\right) = \frac{\partial^p C\left(F_1(\theta_1), \dots, F_p(\theta_p)\right)}{\partial F_1(\theta_1), \dots, \partial F_p(\theta_p)} \tag{3}$$

The corresponding joint PDF can be expressed as the product of copula density $c(\cdot)$ and marginal PDFs with the form,

$$f(\theta_1, \dots, \theta_p) = c\left(F_1(\theta_1), \dots, F_p(\theta_p)\right) \prod_{i=1}^{p} f_i(\theta_i) \tag{4}$$

### 2.2 Variational Inference

Assuming $X$ is a set of observable variables and $\Theta$ is a set of latent variables in a Bayesian model, posterior distribution can be represented by Bayesian formula,

$$p(\Theta|X) = \frac{p(\Theta, X)}{p(X)} = \frac{p(\Theta)p(X|\Theta)}{\int p(\Theta, X)d\Theta} \quad (5)$$

Variational Inference (VI) method uses approximated posterior distribution $q(\Theta)$ to approach the true complex posterior $p(\Theta|X)$ through minimizing the Kullback-Leibler divergence between the approximated distributions and the true one. Compared with point estimation framework, such as Expectation-Maximization (EM), VI gives a complete form of a posterior distribution of latent variables. Solving the best approximated posterior distribution $q^*(\Theta)$ to minimize the Kullback-Leibler divergence,

$$KL\{q(\Theta)||p(\Theta|X)\} = \int q(\Theta) log \frac{q(\Theta)}{p(\Theta|X)} dq(\Theta) \quad (6)$$

However, in practical applications, it is often impossible to solve the integral in the above formula because there is no closed solution to the integral, or because of the exponential computational complexity.

To simplify it, we derive the evidence lower bound (ELBO) $J(q)$ of $\ln p(X)$ from the Kullback-Leibler divergence,

$$J(q) = \ln p(X) - KL\{q(\Theta)||p(\Theta|X)\} = \mathbb{E}_{q(\Theta)}[\ln p(\Theta, X)] - \mathbb{E}_{q(\Theta)}[q(\Theta)] \quad (7)$$

Minimizing the Kullback-Leibler divergence is equivalent to maximizing $J(q)$. The coordinate ascent updates for this optimization problem can be obtained by taking the partial derivatives of $J(q)$, setting the partial derivatives to zero, and solving for the latent variables.

## 2.3 Integrated spline regression

Integrated spline [19], I-spline, is a modification of M-Spline basis function which is attractive for statistician because of its properties, e.g. normalization and the fact that they are defined such that they are positive in $(\zeta_j, \zeta_{j+p})$ and 0 elsewhere. For $p = 1$,

$$M_j^1(x_i) = \begin{cases} \frac{1}{\zeta_{j+1} - \zeta_j}, & \zeta_j \leq x_i < \zeta_{j+1} \\ 0, & otherwise \end{cases} \quad (8)$$

and for $p > 1$, we have,

$$M_j^p(x_i) = \frac{p[(x_i - \zeta_j)M_j^{p-1}(x_i) + (\zeta_{j+p} - x_i)M_{j+1}^{p-1}(x_i)]}{(p-1)(\zeta_{j+p} - \zeta_j)} \quad (9)$$

Using M-Spline basis function, I-Spline can be defined as,

$$I_j^p(x_i) = \int_L^{x_i} M_j^p(u) du \quad (10)$$

Compared with traditional spline basis function, such as B-Spline and truncated spline, if $\beta_j, j = 0,1,2,\ldots,K+p$ are all non-negative values, then I-Spline can produce monotone splines which is in accordance with the characteristic of power curve. For $\zeta_l \leq x_i \leq \zeta_{l+1}$, we can rewrite $I_j^p(x_i)$ in the following form,

$$I_j^p(x_i) = \begin{cases} 0, & j > l \\ \sum_{m=j}^{l}(\zeta_{m+p+1} - \zeta_m)\dfrac{M_m^{p+1}(x_i)}{(p+1)}, & l-p+1 \leq j \leq l \\ 1, & j < l-p+1 \end{cases} \quad (11)$$

Given specific knots [ $\zeta_1, \zeta_2, \ldots, \zeta_K$ ] in $[L, U]$, the I-spline model can be represented as,

$$f(x_i) = \sum_{j=0}^{K+p} \beta_j I_j^p(x_i) \quad (12)$$

where $p$ is the degree of I-Sline basis.

## 3. Methodology

### 3.1 Copula-based Variational Inference

Copula-based variational inference means that copula is regarded as an inference engine for full posterior approximation. All the unknown latent variables in the user-specified hierarchical model are encapsulated into a vector $\boldsymbol{\Theta}$, and the optimal variational approximation $\boldsymbol{q_C(\Theta)}$ to the true posterior $\boldsymbol{p(\Theta|X)}$ is found under the Sklar's representation. This approach provides users with full modeling freedom and does not require conditional conjugacy between latent variables; thus the approach is applicable to general models.

According to Sklar's therom, the CDF of posterior joint distribution in Variational Inference framework can be described with univariate marginal distributions $F_i(\theta_i), i = 1, \ldots, p$ and a tractable copula family $C \in \mathbb{C}$ which describes the dependence structures between variables. We assume $F(\cdot)$ and $C(\cdot)$ to be differentiable, the variational proposal can be constructed as,

$$q_C(\boldsymbol{\Theta}) = c_\Upsilon\left(F_1(\theta_1), \ldots, F_p(\theta_p)\right) \prod_{i=1}^{p} f_i(\theta_i) \quad (13)$$

where $c_\Upsilon$ is the copula density with parameter $\Upsilon$; $f_i(\theta_i)$ and $F_i(\theta_i)$ are the PDF and CDF for $i$-th latent variable respectively. Let the true posterior $p(\boldsymbol{\Theta}|X)$ in Sklar's representation be $p(\boldsymbol{\Theta}|X) = c^*\left(F_1^*(\theta_1), \ldots, F_p^*(\theta_p)\right) \prod_{i=1}^{p} f_i^*(\theta_i)$, where $c^*(\cdot)$ and $\{f_i^*(\theta_i)\}_{i=1:p}$ are the true underlying copula density and marginal posterior densities, respectively. Then, the KL divergence decomposes into copula terms and PDF terms. The derivations are provided in Appendix A.

$$KL\{q_C(\Theta)||p(\Theta|X)\} = KL\{c[F(\Theta)]||c^*[F^*(\Theta)]\} + \sum_{i=1}^{p} KL\{f_i(\theta_i)||f_i^*(\theta_i)\} \tag{14}$$

The ELBO of copula variational approximation is,

$$J(q) = \int c[F(\Theta)] \times \prod_{i=1}^{p} f_i(\theta_i) \ln p(X, \Theta) d\Theta - \int c[F(\Theta)] \ln c[F(\Theta)] dF(\Theta)$$
$$+ \sum_{i=1}^{p} \int f_i(\theta_i) \ln f_i(\theta_i) d\theta_i \tag{15}$$

However, directly optimizing the ELBO in the above formula often leads to a non-trivia variational calculus problem. For computational convenience, we present equivalent proposal constructions based on reparameterization.

We incorporate auxiliary variables $Z$ by exploiting the latent variable representation: $Z$ follows the joint distribution defined by copula $C$; $g(\cdot)$ is the transformation function by mapping domain $[0,1]$ of $F_i(\theta_i)$ to the desired domain of copula density function $c_\Psi$ with parameters $\Psi$. Therefore, $\theta_i$ can be represented as: $\theta_i = F_i^{-1}(g(z_i))$, $Z \sim c_\Psi$. We use the example of Gaussian copula and Student-t copula to illustrate the transformation. For Gaussian copula, $g(\cdot) = \Phi^{-1}(\cdot), [0,1] \to [-\infty, +\infty], c_\Psi = N_p(\mathbf{0}, \Sigma)$ where $\Phi^{-1}(\cdot)$ is the inverse function of the standard normal distribution and $\Sigma$ is covariance matrix; for Student-t copula, $g(\cdot) = T_v^{-1}(\cdot), [0,1] \to [-\infty, +\infty], c_\Psi = T_p(v, \Sigma)$ where $T_v^{-1}(\cdot)$ is the inverse function of univariate Student-t distribution with freedom $v$. $g(\cdot)$ is monotonic in most of the cases to guarantee one to one mapping of $\theta_i$ and $z_i$., $g(\cdot)$ is non-decreasing for Gaussian copula and Student-t copula, while $g(\cdot)$ is monotone decreasing for Archimedean copulas.

Letting $h(\cdot) = F_i^{-1}(g(\cdot))$, $\theta_i = h(z_i)$, $h(z_i)$ is a monotone function and is continuous on its domain $\mathcal{Z}$, and $h^{-1}(\cdot)$ has a continuous derivative. According to the distribution transformation theorem, $q_{VC}(\Theta)$ can be represented as,

$$q_{VC}(\Theta) = q_c(h^{-1}(\Theta); \Psi) [\prod_{i=1}^{p} |\frac{d}{d\theta_i} h_i^{-1}(\theta_i)|] \tag{16}$$

The ELBO of copula variational approximation can be reparametrized by $q_{VC}(\Theta)$. The derivations are provided in Appendix B. If $h(\cdot)$ is monotonous, then,

$$J(q) = \begin{cases} \mathbb{E}_{q_c(\mathbf{Z};\mathbf{\Psi})}\left[logp(h(\mathbf{Z}),\mathbf{X}) - logq_c(\mathbf{Z};\mathbf{\Psi}) + \sum_{i=1}^{p} \log h'_i(z_i)\right] & h' > 0 \\ (-1)^p \mathbb{E}_{q_c(\mathbf{Z};\mathbf{\Psi})}\left[logp(h(\mathbf{Z}),\mathbf{X}) - logq_c(\mathbf{Z};\mathbf{\Psi}) + \sum_{i=1}^{p} \log |h'_i(z_i)|\right] & h' < 0 \end{cases} \quad (17)$$

After reparameterization, $J(q)$ is more tractable by setting an approximate form of $q_c(\mathbf{Z};\mathbf{\Psi})$; $h(\cdot)$ serves as a monotonous transformation to allow latent random variable $\mathbf{\Theta}$ to follow more complicated distribution.

## 3.2 Sequential Incomplete Profile Construction

In this section, we construct profiles for WT during certain period (e.g. 3 days). As we mentioned above, profiles may contain incomplete range of wind speed due to the intermittence and volatility of wind which makes it difficult to compare WT performance in different periods. To solve it, we introduce the concept "prior" in Bayesian learning. The posterior distribution of latent variables in former period is used as the prior distribution of that in latter period. The prior distribution contains marginal distribution and covariance matrix of latent variables. In this way, we obtain the latest distribution of latent variables inferenced by copula-based variational inference (CVI).

Given the training samples $\{x_i, y_i\}$ of size $N$ in each period, the power curve model can be represented as $y_i = f(x_i) + e_i$, where $f$ is the profile function and $e_i$ is the error term. In this paper, we construct the I-Spline regression model inferenced by CVI. The main reason to use I-Spline is that: I-Spline can guarantee the monotonicity of the input; therefore, it can combine with approximate hierarchical model and CVI to construct monotone profile with latent variable distribution indicating WT performance.

Giving the training samples of size $N$, the I-spline regression model can be written as the matrix form,

$$\mathbf{Y} = \mathbf{Z}\boldsymbol{\beta} + \mathbf{e} \quad (18)$$

where $\mathbf{Y}$ represents wind power vector, $\mathbf{Y} \in \mathcal{R}^N$; $\mathbf{Z}$ contains model input matrix generated from wind speed vector, $\mathbf{Z} \in \mathcal{R}^{N\times p}$; $\boldsymbol{\beta}$ is model coefficient and indicates WT performance, $\boldsymbol{\beta} \in \mathcal{R}^p$; $\mathbf{e}$ is error term, $\mathbf{e} \in \mathcal{R}^N$.

We use proposed CVI with Gaussian copula $c_G(\cdot)$ to model dependence and Log-Normal distribution to model marginal distribution of latent variables. A Gaussian copula function with $p \times p$ correlation matrix $\Sigma$ is defined as,

$$C_G(F(\Theta)|\Upsilon) = \Phi_p\left(\Phi^{-1}(F(\theta_1)), \ldots, \Phi^{-1}\left(F(\theta_p)\right)\right): [0,1]^p \to [0,1] \tag{19}$$

where $\Phi(\cdot)$ is the CDF of $\mathcal{N}(0,1)$, and $\Phi_p(\cdot|\Sigma)$ is the CDF of $\mathcal{N}_p(\mathbf{0},\Sigma)$. The Gaussian copula density is,

$$c_G(F(\Theta)|\Sigma) = \frac{1}{\sqrt{\Sigma}} \exp\left\{-\frac{\mathbf{z}^T(\Sigma^{-1} - I_p)\mathbf{z}}{2}\right\} \tag{20}$$

where $\mathbf{z} = [\Phi^{-1}(F(\theta_1)), \ldots, \Phi^{-1}(F(\theta_p))]^T$. The prior distribution on regression coefficient $\boldsymbol{\beta}_j$ at period $t$ is assumed to be Log-Normal distribution with mean $u_{j,t-1}$ and variance $\sigma_{j,t-1}$.

$$p(\beta_{j,t}) = \mathcal{LN}(u_{j,t-1}, \sigma_{j,t-1}^2) \tag{21}$$

The error term $e_{i,t}$ is assumed to obey a Gaussian distribution with mean 0 and variance $\tau_i^{-1}$, which is given a Gamma prior distribution to complete the Bayesian model,

$$\begin{aligned} p(\beta_{j,t}) &= \mathcal{LN}(u_{j,t-1}, \sigma_{j,t-1}^2) \\ p(e_{i,t}|\tau_{i,t}) &= \mathcal{N}(e_{i,t}|0, \tau_{i,t}^{-1}), \qquad p(\tau_{i,t}) = \mathcal{G}(\tau_{i,t}|a_0, b_0) \end{aligned} \tag{22}$$

The corresponding parameters can be represented as $\Theta_t = \{\boldsymbol{\beta}_t, \boldsymbol{\tau}_t\}$, where $\boldsymbol{\beta}_t = \{\beta_{1,t}, \beta_{2,t}, \ldots, \beta_{p,t}\}, \boldsymbol{\tau}_t = \{\tau_{1,t}, \tau_{2,t}, \ldots, \tau_{N,t}\}$. The posterior distribution of $\Theta_t$ can be constructed via a Sklar's theorem:

$$q(\Theta_t) = c_G(F(\Theta_t)|\Sigma) \prod_{j=1}^{p} LN(\boldsymbol{\beta}_{j,t}; u_{j,t}, \sigma_{j,t}^2) \prod_{i=1}^{N} LN(\boldsymbol{\tau}_{i,t}; c_{i,t}, d_{i,t}^2) \tag{23}$$

where $\Theta_t = \{\beta_{1,t}, \beta_{2,t}, \ldots, \beta_{p,t}, \tau_{1,t}, \tau_{2,t}, \ldots, \tau_{N,t}\}$; the mean and variance of $\boldsymbol{\tau}_{i,t}$ are represented as $c_{i,t}$ and $d_{i,t}^2$ respectively to avoid confoundment. However, it's not easy to directly solve $q(\Theta_t)$ via CVI because $\Sigma$ is a positive semi-definite matrix with ones on the diagonal and off-diagonal elements between $[-1,1]$. For Gaussian copula, we adopt the parameter expansion technique [20, 21] mentioned in paper [22],

$$\tilde{z}_i = t_i^{-1}(z_i) = u_i + \sigma_{ii} z_i, \qquad \tilde{\mathbf{z}} \sim \mathcal{N}_p(\boldsymbol{\mu}, D\Sigma D^T) \tag{24}$$

where $D = [diag(\sigma_{ii})]_{i=1:p}$. Therefore, $\theta_i = F_i^{-1}(\Phi(t(\tilde{z}_i))) \coloneqq r_i(\tilde{z}_i)$, where $r_i(\cdot) = F_i^{-1} \circ (\Phi \circ t(\cdot))$ is monotonic increasing function.

To inference the distribution of interested latent variables, $J(q)$ is represented as,

$$J(q) \propto \left(a_0 + \frac{1}{2}\right)\sum_{i=1}^{N} <\ln\tau_{i,t}> - \frac{1}{2}\sum_{i=1}^{N} <\tau_{i,t}><(y_{i,t} - \mathbf{Z}_{i,t}^T\boldsymbol{\beta}_t)^2>$$
$$-b_0\sum_{i=1}^{N}<\tau_{i,t}> - \sum_{j=1}^{p}\frac{<(\ln\beta_j - u_{j,t-1})^2>}{2\sigma_{j,t-1}^2} + \frac{1}{2}\ln|\boldsymbol{\Sigma}_{\beta,t}|$$
(25)

where $<\cdot>$ means the expectation value of random variable; $\boldsymbol{\Sigma}_{\beta,t}$ is a $p \times p$ correlation matrix of $\boldsymbol{\beta}$ and is an inherent symmetric matrix.

$$\boldsymbol{\Sigma}_{\beta,t} = \begin{pmatrix} \sigma_{1,t}^2 & \sigma_{1,2,t} & \cdots & \sigma_{1,p-1,t} & \sigma_{1,p,t} \\ \sigma_{2,1,t} & & & & \sigma_{2,p,t} \\ \vdots & & \ddots & & \vdots \\ \sigma_{p-1,1,t} & & & & \sigma_{p-1,p,t} \\ \sigma_{p,1,t} & \sigma_{p,2,t} & \cdots & \sigma_{p,p-1,t} & \sigma_{p,t}^2 \end{pmatrix}$$

The coordinate ascent updates for updating $J(q)$ can be obtained by taking the partial derivatives of $J(q)$, setting the partial derivatives to zero, and solving for the parameters $u_{j,t}, \sigma_{j,t}^2, \sigma_{2,1,t}, c_{i,t}, d_{i,t}^2$. This yields coordinate updates:

$$\frac{\partial J(q)}{\partial u_{j,t}} = \left(\sum_{i=1}^{N} B_{i,j,t}(y_{i,t} - C_{i,j,t})\right)\exp\{u_{j,t} + \frac{\sigma_{j,t}^2}{2}\} - D_{j,t}\exp\{2u_{j,t} + 2\sigma_{j,t}^2\} - \frac{u_{j,t} - u_{j,t-1}}{\sigma_{j,t-1}^2} = 0$$

$$\frac{\partial J(q)}{\partial \sigma_{j,t}^2} = \left(\sum_{i=1}^{N} B_{i,j,t}(y_{i,t} - C_{i,j,t})\right)\exp\{u_{j,t} + \frac{\sigma_{j,t}^2}{2}\} - 2D_{j,t}\exp\{2u_{j,t} + 2\sigma_{j,t}^2\} - \frac{1}{\sigma_{j,t-1}^2}$$
$$+ \frac{A_{j,j,t}}{\sigma_{j,t}^2 A_{j,j,t} + \sum_{k \neq j} A_{j,k,t}\sigma_{j,k,t}} = 0$$

$$\frac{\partial J(q)}{\partial \sigma_{j,k,t}} = -H_{j,k,t}I_{j,k,t}\exp\{\sigma_{j,k,t}\} + \frac{1}{2}\frac{2a_{1,j,k,t}\sigma_{j,k,t} + a_{2,j,k,t}}{a_{1,j,k,t}(\sigma_{j,k,t})^2 + a_{2,j,k,t}\sigma_{j,k,t} + a_{3,j,k,t}} = 0$$

$$\frac{\partial J(q)}{\partial c_{i,t}} = \exp\{c_{i,t} + \frac{d_{i,t}^2}{2}\}\left(b_0 + \frac{<(y_{i,t} - \mathbf{Z}_{i,t}^T\boldsymbol{\beta})^2>}{2}\right) - \left(a_0 + \frac{1}{2}\right) = 0$$

$$\frac{\partial J(q)}{\partial d_{i,t}^2} = \exp\{c_{i,t} + \frac{d_{i,t}^2}{2}\}\left(b_0 + \frac{<(y_{i,t} - \mathbf{Z}_{i,t}^T\boldsymbol{\beta})^2>}{2}\right) - \frac{1}{2d_{i,t}^2} = 0$$

where $A_{j,k,t}$ is the algebraic complement of element $\boldsymbol{\Sigma}_{\beta,t}[j,k]$ and it equals $(-1)^{j+k}M_{j,k,t}$, $M_{j,k,t}$ is the determinant after removing the $j$-th row and $k$-th column of $\boldsymbol{\Sigma}_{\beta,t}$; $B_{i,j,t} = Z_{i,j,t}<\tau_{i,t}>$, $C_{i,j,t} = \sum_{k \neq j} Z_{i,k,t}e^{\sigma_{j,k,t}}<\boldsymbol{\beta}_{k,t}>$, $D_{j,t} = \sum_{i=1}^{N} Z_{i,j,t}^2<\tau_{i,t}>$, $H_{j,k,t} = \sum_{i=1}^{N}<\tau_{i,t}>Z_{i,j,t}Z_{i,k,t}$; $I_{j,k,t} = \exp\{u_{j,t} + u_{k,t} + \frac{\sigma_{j,t}^2 + \sigma_{k,t}^2}{2}\}$; $a_{1,t}, a_{2,t}$ and $a_{3,t}$ are linear combination of elements in $\boldsymbol{\Sigma}_{\beta,t}$ except $\sigma_{j,k,t}$ and $\sigma_{k,j,t}$, $|\boldsymbol{\Sigma}_{\beta,t}| = a_{1,j,k,t}(\sigma_{j,k,t})^2 + a_{2,j,k,t}\sigma_{j,k,t} + a_{3,j,k,t}$.

The partial derivative equations contain exponential terms and it difficult to calculate the explicit solution. Therefore, Newton-Raphson method is used to calculate the nonlinear equation.

The key of Newton-Raphson method is to set an approximate initial guess value. The sequential updating property of our modeling provides convenience for our problem because the prior distribution value can be used as a reliable initial guess. Besides, we set a boundary in the solving process to get the feasible solution. The updating scheme of N-R method is defined as follows. If the guess changes from one step to the next without exceeding a threshold τ, then the algorithm stops and confirms that the latest guess is close enough. If we reach a certain number of guesses but still do not reach the threshold, then we give up continuing to guess. Actually, because the initial value is close to the final solution, we can always get the satisfied solution for the equation.

In summary, based on the principle of sequential prior-posterior transformation and CVI, the sequential incomplete profile construction algorithm can be summarized in Algorithm 1. In the real profile monitoring process, to track the latest characteristics of producing wind energy, the posterior distribution of model coefficients for the former data segment is used as the prior distribution of that for the latter data segment. Because samples in different data segments are different, posterior distributions of error term $[c_t, d_t^2]_{i=1:N}$ don't participate in the updating process.

---

**Algorithm 1**

---

**Input**: $maxiter, \mu_{j=1:p,0}, \sigma_{j=1:p,0}, a_0, b_0$.

**Output**: $u_{j=1:p,t}, \sigma^2_{j=1:p,t}; t = 1, \ldots, T$

Initialize: $\boldsymbol{\beta}, \boldsymbol{e}$.

**for** $t = 1, t < T$ **do**

    generating model inputs for data segment #$t$ according to I-spline function

    **for** $epoch = 1, epoch < maxiter$ **do**

        $[\boldsymbol{u}_t, \boldsymbol{\sigma}_t^2]_{j=1:p}, \sigma_{j,k,t}(j \neq k)$ are optimized

        $[\boldsymbol{c}_t, \boldsymbol{d}_t^2]_{i=1:N}$ are optimized

    **end**

    updating: $u_{j=1:,p,t-1}, \sigma^2_{j=1:p,t-1}$ using latest calculated $[\boldsymbol{u}_t, \boldsymbol{\sigma}_t^2]_{j=1:p}$

**End**

**return** $u_{j=1:p,t}, \sigma^2_{j=1:p,t}; t = 1, \ldots, T$

---

## 3.3 Directional performance degradation detection

The performance degradation process which means that power generation efficiency declines significantly can be step by step during the operation process or broken out suddenly due to rugged environment. Unlike traditional change point detection research where the anomaly is sparse, the degradation process of wind turbines has certain direction with the movement of the whole power curve, as shown in Fig. 2. Therefore, we designed a directional online performance degradation detection algorithm for wind turbines combined with the former sequential incomplete power curve construction method.

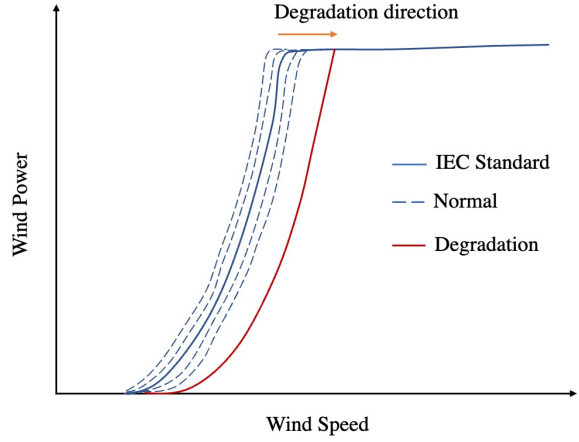

Fig 2. Illustration of power curve with performance degradation

For normal power curve with no degradation, we assume,

$$Y_t = Z_t \beta_t + e_t, \qquad t \leq \tau$$

If the wind turbine occurs performance degradation, then the wind power efficiency decreases, which means that the power curve shifts to certain direction. The formula of the degradation profile can be represented as,

$$Y_t = Z_t \beta_t - Z_t \xi_t^2 + e_t, \qquad t > \tau$$

Combining the above two equations, we can define the change point model as this: $\xi_t^2 = \mathbf{0}$ for $t \leq \tau$ and $\xi_t^2 \neq \mathbf{0}$ for $t > \tau$. However, because wind turbine performance degradation comes with the movement of whole power curve, not sparse anomalies, it is not suitable to solve the hypothesis because the calculation process causes a non-trivial variational calculus problem. In order to solve this problem, we propose a method to compare current power curve $\mathcal{P}_t$ with the IEC standard power curve $\mathcal{P}_0$. As model coefficients of the proposed monotonic profile are aligned with power generation efficiency given specific wind speed, performance degradation indicates that the mean of the posterior distribution of model coefficients declines. We define a new hypothesis as,

$$H_0: \boldsymbol{\beta}_t \sim \mathcal{LN}(\boldsymbol{u_0}, \boldsymbol{\Sigma_0}), \quad H_1: \boldsymbol{\beta}_t \sim \mathcal{LN}(\boldsymbol{u_0} - \boldsymbol{d}, \boldsymbol{\Sigma_1})$$

where $\boldsymbol{d}$ is a non-negative vector and we assume that the reduced degree of each model coefficient is the same, i.e., the amplitude of wind power generation decline is the same during different wind speed scope when wind turbine occurs performance degradation.

To decide whether performance degradation happens to the system, we need to quantify the evidence for hypothesis $H_1$ to $H_0$. This task is different to previous methods mentioned in paper [23] as the monitored object is posterior distribution of model coefficients rather than sample values. To solve it, we proposed KL-divergence factor *KLF* to measure the evidence, that is,

$$KLF = \frac{p(\boldsymbol{\beta}_t|H_0)}{p(\boldsymbol{\beta}_t|H_1)} \propto \frac{KL(\mathcal{LN}(\boldsymbol{u}_\beta, \boldsymbol{\Sigma}_\beta)||\mathcal{LN}(\boldsymbol{u}_0, \boldsymbol{\Sigma}_0))}{KL(\mathcal{LN}(\boldsymbol{u}_\beta, \boldsymbol{\Sigma}_\beta)||\mathcal{LN}(\boldsymbol{u}_0 - \boldsymbol{d}, \boldsymbol{\Sigma}_1))} \quad (26)$$

*KLF* is coherent in our sequential change detection framework. Considering the latest posteriors of parameters are updated sequentially as new observation comes, they can compare with IEC standard distribution and pre-set alarm distribution to test whether wind turbine occurs performance degradation at present.

After some simplifications, *KLF* can be derived as the final detection statistic $\Lambda_n$, that is,

$$\Lambda_n = \frac{(\boldsymbol{\mu}_\beta - \boldsymbol{\mu}_0)^T \boldsymbol{\Sigma}_0^{-1}(\boldsymbol{\mu}_\beta - \boldsymbol{\mu}_0) + \log\left(\frac{|\boldsymbol{\Sigma}_0|}{|\boldsymbol{\Sigma}_\beta|}\right) + tr(\boldsymbol{\Sigma}_0^{-1}\boldsymbol{\Sigma}_\beta) - K}{\left(\boldsymbol{\mu}_\beta - (\boldsymbol{\mu}_0 - \boldsymbol{d})\right)^T \boldsymbol{\Sigma}_1^{-1}\left(\boldsymbol{\mu}_\beta - (\boldsymbol{\mu}_0 - \boldsymbol{d})\right) + \log\left(\frac{|\boldsymbol{\Sigma}_1|}{|\boldsymbol{\Sigma}_\beta|}\right) + tr(\boldsymbol{\Sigma}_1^{-1}\boldsymbol{\Sigma}_\beta) - K} \quad (27)$$

The derivations are provided in Appendix D. We can set a detection threshold $h$ according to a pre-specific confidence level (false alarm rate), and define that if $\Lambda_n > h$, the test statistic triggers an abnormal alarm.

There are two hyperparameters we need to decide before running the directional performance degradation detection algorithm: $\boldsymbol{d}$ and $h$. $\boldsymbol{d}$ can be determined according to the tolerance of the power generation efficiency decline, for example, $0.1\boldsymbol{u}_0$ ($\boldsymbol{u}_0 > \boldsymbol{0}$). After specifying the value of $\boldsymbol{d}$, the detection threshold $h$ is determined by controlling a pre-specific false alarm rate, or equivalently setting the in-control ARL to a specific large number, for example, 200. An empirical method can be used to find the threshold based on empirical distribution of $\Lambda_n$. The most commonly used method is based on Monte Carlo simulation, and the detailed algorithm can be referred to paper [24].

# 4. Experiments

## 4.1 Data description

In this research, we use the penmanshiel wind farm dataset which locates in the east of the United Kingdom [25]. There are 14 wind turbines in this wind farm. The location of the wind farm and relative location of wind turbines are shown in Fig. 3 and Fig. 4 respectively. The construction of this wind farm is finished at the second half of 2016. We have the operation data of this wind farm including wind speed, wind direction, ambient temperature and wind power output from 2017/01/01 to 2021/06/30 with time resolution being 10 min.

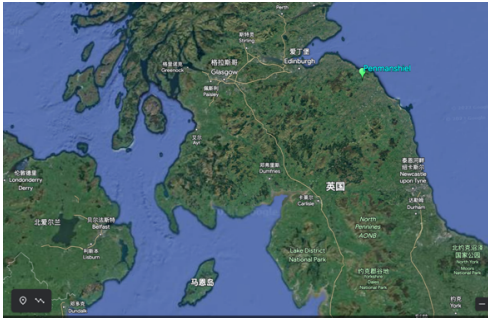 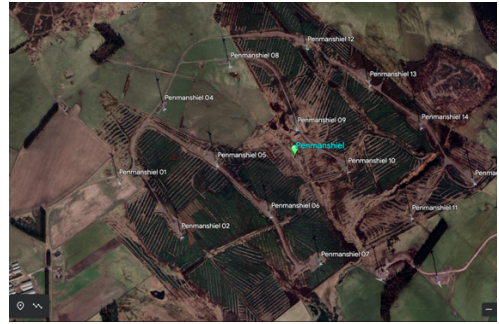

Fig 3. Location of the wind farm　　　　Fig 4. Relative location of wind turbines

After analyzing SCADA data, there exists performance degradation for Penmanshiel 02 and Penmanshiel 08 in year 2021 and there is no performance degradation for other wind turbines. It can be seen in Fig. 5 that Penmanshiel 02 and Penmanshiel 08 occurs performance degradation with mean decrease and variance increase from 2021-01-01 to 2021-07-01 while the power curves of Penmanshiel 02 remained steady during different time periods.

There are plenty of outliers in the SCADA data which are caused by shut down, limit power state or sensor recording error. These outliers need to be removed, otherwise, they may compromise the accurate depiction of the power curves which reflects the current wind power generation ability. In this paper, unlike traditional outlier removing method with high complexity, we design a new outlier removing method. SCADA data for a specific wind turbine at time $t$ is represented as $X_t$, which contains information about wind speed, wind power. $X_t$ is represented as $X_t = [v_t, p_t]$. For each power curve, we generated 80 wind speed bins by traversing wind speed from 5 m/s to 13m/s with step 0.1m/s. Wind speed bin $B_i$ can be represented as,

$$B_i = \{X_t : v_t \in [v_i, v_i + 0.1)\}$$

$$v_i = 5 + 0.1 * (i - 1), \quad i = 1, 2, \ldots, 80$$

where $v_i$ is the traversed wind speed; $v_t$ is the wind speed at time $t$. Then, in each wind speed bin, wind power samples are used to fit Gaussian distribution $f(x) = \frac{1}{\sqrt{2\pi}\sigma} e^{-\frac{(x-\mu)^2}{2\sigma^2}}$. The 1% quantile of $f(x)$ can be obtained and used as the threshold to identify outliers. If $p_t$ exceeds the calculated threshold, the $X_t$ is regarded as outlier and removed.

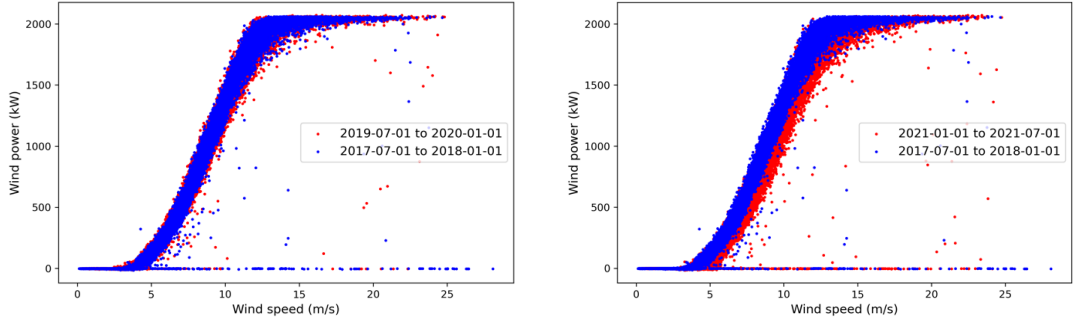

(a) Represented power curves for Penmanshiel 02

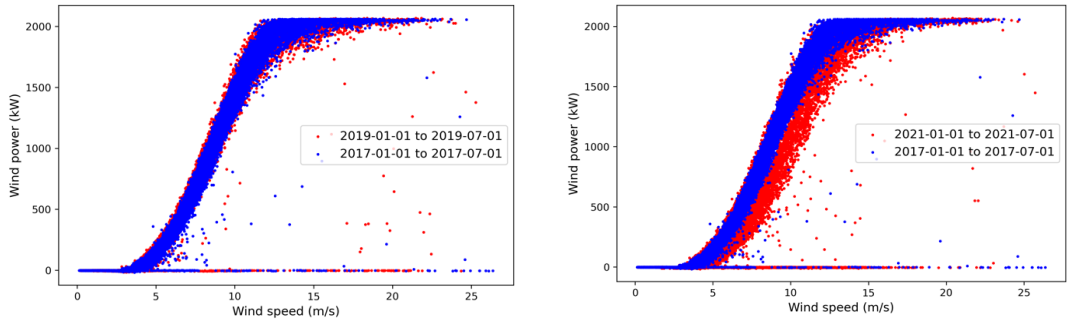

(b) Represented power curves for Penmanshiel 08

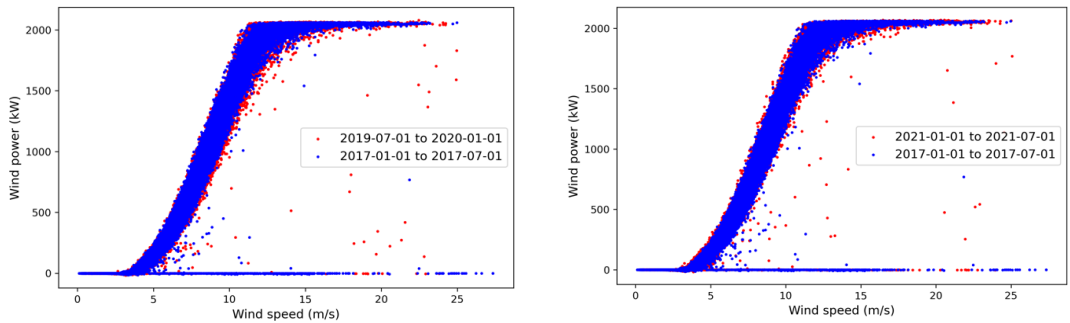

(c) Represented power curves for Penmanshiel 15

Fig 5. Represented power curves of three wind turbines

## 4.2 Sequential incomplete power curve modeling

### 4.2.1 Experiment setting

The time series data of wind speed and wind power output are divided into data segments with equal time length to facilitate later profile modeling and degradation detection. Rolling window method [26] is often used to do this. The SCADA records with length $L$ are cut into segments with equal length $N_w$, and each data segment will be updated with $N_u$ new data. The smaller $N_u$ is, the higher the degree of overlap between consecutive data segments. In this paper, $N_w = 500$ or $1000$ which corresponds to short-term and medium-term performance degradation detection respectively, and $N_u = N_w/2$. Incomplete power curve monitoring task mainly occurs in these two periods.

Besides the values of $N_w$ and $N_u$, knots of I-spline basis, value of $a_0, b_0$, maximum number of iterations $maxiter$ need to specify. In this paper, knots are set by taking values between cut-in wind speed $v_c$ ($3m/s$) and rated wind speed $v_r$ ($14m/s$) with equal step size ($1m/s$). The order of I-Spline is set 3. As to hyperparameters $\{a_0, b_0\}$, they are set as 0.1. $maxiter$ is set 30.

To test the effectiveness of proposed method (CVI), multivariate Gaussian proposal (MGR) [27, 28], Weibull cumulative distribution function (WCDF)-based power curve [15] and Gaussian process regression (GPR) [11] are employed as benchmark models. DL-based models such as long-short term memory (LSTM) are not used because these models have complex parametric forms and it's difficult to design monitoring scheme for them. Multivariate Gaussian proposal uses Gaussian distribution as marginal distribution and Gaussian copula to describe dependence structure. The detail of MGR framework is attached in Appendix C. The equation of WCDF is stepwise: $P = 1 - e^{-(x/c)^k}$ when $v_c < v < v_r$; $P = 0$ when $v < v_c$; $P = 1$ when $v > v_c$. For GPR, radial basis function (RBF) is used as kernel function.

Two datasets including the operation data of Penmanshiel 02 and Penmanshiel 08 are used for latter analysis. Three error indicators, mean absolute error (MAE) [29], root mean squared error (RMSE) [30] and mean absolute percentage error (MAPE) [31], are used to measure the performances proposed method and different baselines. To make a fair model comparison, all models are trained with the same data samples, and they all have only one input (wind speed) and one output (wind power).

### 4.2.2 WTPC modeling

According to the principles of sequential power curve modeling, that is the posterior distribution of model coefficients for the former data segment is used as the prior distribution of that for the latter data segment, a good starting posterior is necessary for latter sequential power curve updating. When wind turbine is just put into use, wind power generation efficiency is high and there is generally no performance degradation. Therefore, we need to construct a good starting posterior distribution using adequate data samples which includes complete scopes of wind speed before subsequent sequential updating and degradation detection. The starting posterior distribution is also used as the parameter distribution of IEC standard power curve $\mathcal{P}_0$ to construct hypothesis $H_0$.

We use the data of Penmanshiel 08 illustrate the construction of wind power curve. $N_w$ is set 500 and $N_u$ is set 250. Each period contains data in approximately half-week. To obtain a good starting posterior distribution, the first 1000 data samples with whole wind speed scope are used to infer the posterior distribution of model coefficients. It can be seen in Fig. 6 that the fitted wind power curve is aligned with raw values which reflects the true operation state of WT Penmanshiel 08.

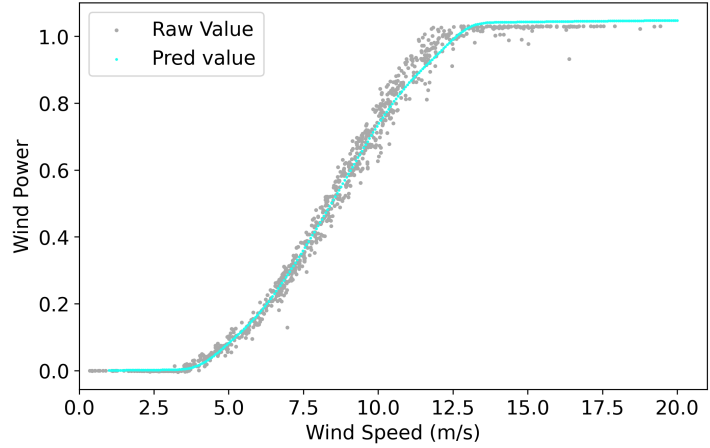

Fig 6. Fitting result of the starting data segment for Penmanshiel 08

After obtaining the posterior distribution of the starting data segment, we use Algorithm 1 to update model coefficients sequentially. Partial results for proposed CVI and other three baselines for Penmanshiel 08 are shown in Fig. 7. It can be seen that:

(i) Proposed CVI fits the data samples well and constructs reliable power curves whether the

sample size is sufficient or not.

(ii) MGR responds to characteristics of data samples slowly when updating times are large, when can be seen in Fig. 7(d). This is because the posterior variance of model coefficients under MGR framework is severely under-estimated which we will discuss further latter.

(iii) When data samples are sufficient, GPR and WCDF construct reliable power curves. However, the power curves of GPR and WCDF deviates from normal trend if data samples are insufficient as seen in Fig. 7(b) and Fig. 7(c). Although these two methods may fit existing data samples well, they may induce false alarms of performance degradation.

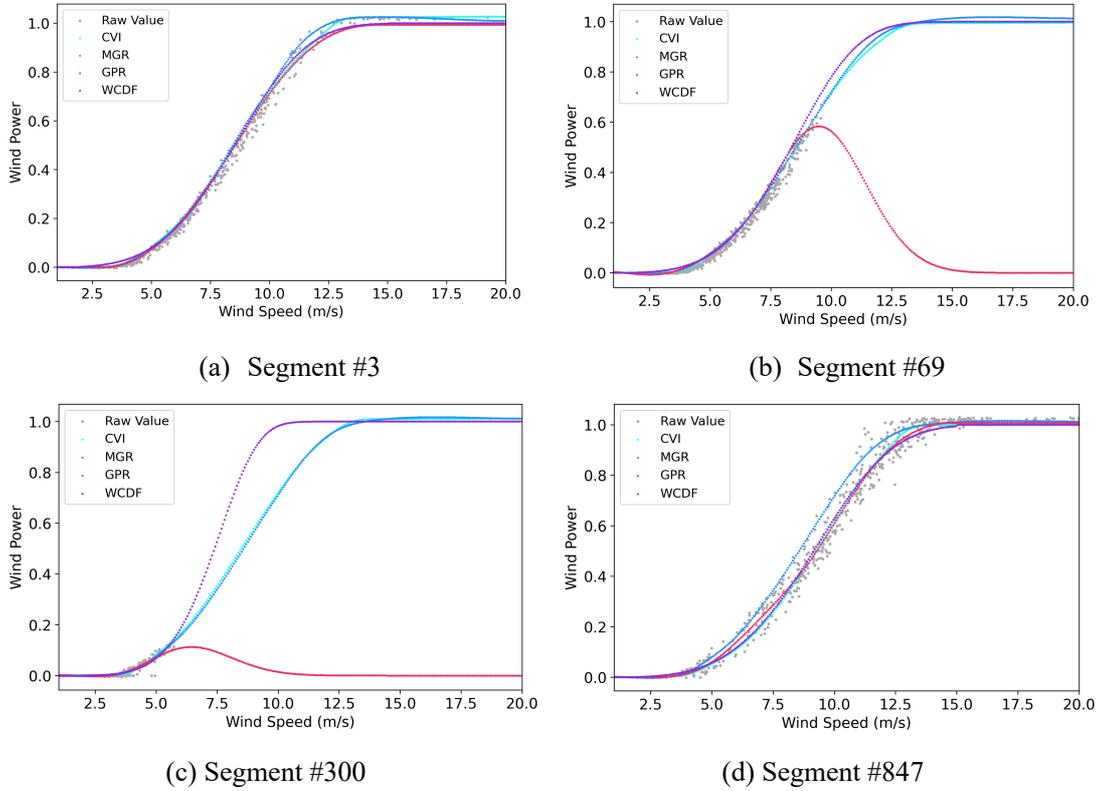

(a) Segment #3  (b) Segment #69

(c) Segment #300  (d) Segment #847

Fig 7. Power curve construction based on 4 methods when $N_w = 500$ for Penmanshiel 08

To illustrate the difference between CVI and MGR, we plot the probability density function of these two methods for the 300-th segment of Penmanshiel 08 as shown in Fig. 8. It can be seen in Fig. 8 that the posterior variance of lognormal distribution inferenced by CVI is significantly larger than the normal distribution inferenced by MGR. Therefore, CVI can achieve better balance between the information of prior and latest data samples. However, MGR is insensitive to latest data samples with updating because its confidence in parameter estimation is huge, i.e., variance is small. In conclusion, CVI outperforms MGR in sequential updating problems.

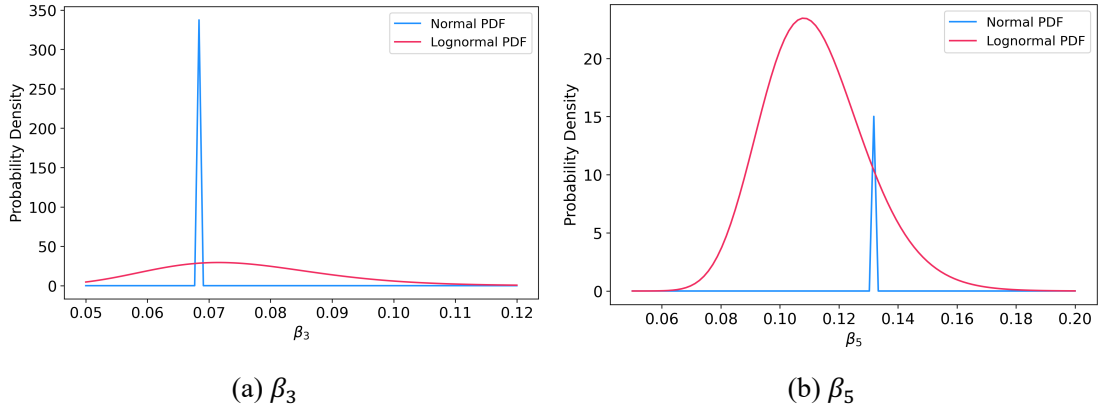

(a) $\beta_3$          (b) $\beta_5$

Fig 8. Posterior PDF of partial model coefficient for the 300-th Segment of Penmanshiel 08

In two datasets, compared with three benchmarks, the prediction performance of proposed method is presented when $N_w = 500, 1000$ respectively as shown in Table 1. From Table 1, CVI generally outperforms other three WTPC models. In terms of RMSE and MAE for prediction results, CVI is the lowest in both two datasets. For MAPE, CVI is the lowest for Penmanshiel 02 and rank #2 for Penmanshiel 08. The prediction performance of GPR improves when the sample size increases while WCDF and MGR have stable prediction performance in four experiment settings.

Table 1. Comparison of different WTPC models

| $N_w$ | Method | Penmanshiel 02 | | | Penmanshiel 08 | | |
|---|---|---|---|---|---|---|---|
| | | RMSE | MAE | MAPE | RMSE | MAE | MAPE |
| 500 | CVI | **0.0246** | **0.0166** | **0.264** | **0.0248** | **0.0163** | 0.285 |
| | MGR | 0.0301 | 0.0192 | 0.283 | 0.0331 | 0.0196 | 0.323 |
| | GPR | 0.0271 | 0.0171 | 0.654 | 0.0266 | 0.0163 | 0.445 |
| | WCDF | 0.0255 | 0.0187 | 0.275 | 0.0252 | 0.0179 | **0.258** |
| 1000 | CVI | **0.0251** | **0.0167** | 0.267 | **0.0250** | 0.0163 | 0.281 |
| | MGR | 0.0301 | 0.0191 | 0.299 | 0.0331 | 0.0195 | 0.283 |
| | GPR | 0.0262 | 0.0169 | 0.672 | 0.0260 | 0.0163 | 0.289 |
| | WCDF | 0.0266 | 0.0194 | 0.267 | 0.0264 | 0.0187 | **0.256** |

## 4.3 Performance degradation detection analysis

In this section, we first present the results of proposed method to detect wind turbine performance degradation. Then we compare it with several baselines. The evaluation metrics used for performance comparison include precision rate, recall rate and F-beta coefficient [32].

As we know, Penmanshiel 02 and Penmanshiel 08 occurs performance degradation during

2021-01-01~2021-07-01. After seeing the operation manual, Penmanshiel 02 occurs performance degradation from 2021-01-11 16:00:00 to 2021-04-27 17:20:00, which corresponds to data segment #818 and data segment #876 respectively when $N_w = 500$, data segment #409 and data segment #437 respectively when $N_w = 1000$; Penmanshiel 08 occurs performance degradation from 2021-01-30 09:20:00 to 2021-05-10 17:10:00, which corresponds to data segment #837 and data segment #888 respectively when $N_w = 500$, data segment #418 and data segment #444 respectively when $N_w = 1000$.

As we mentioned in Section 5.2.2, we construct a power curve with sufficient data samples and its posterior distribution is used to construct hypothesis $H_0$. Then, we run above-mentioned sequential incomplete power curve modeling algorithm and detect performance degradation by comparing calculated $\Lambda_n$ with $h$. In this task, $\boldsymbol{d}$ is set $0.1\boldsymbol{\mu}_0$, which indicates that if the power generation efficiency of WT declines by 10%, then it is judged with occurring performance degradation. $h$ is set 1 and used as the threshold to detect performance degradation. The degradation detection results of proposed method for Penmanshiel 02 and Penmanshiel 08 are shown in Fig. 9 and Fig. 10 respectively. The proposed method successfully detects performance degradation during 2021-01-01~2021-07-01. It seldom triggers false alarms when WT is in normal operation.

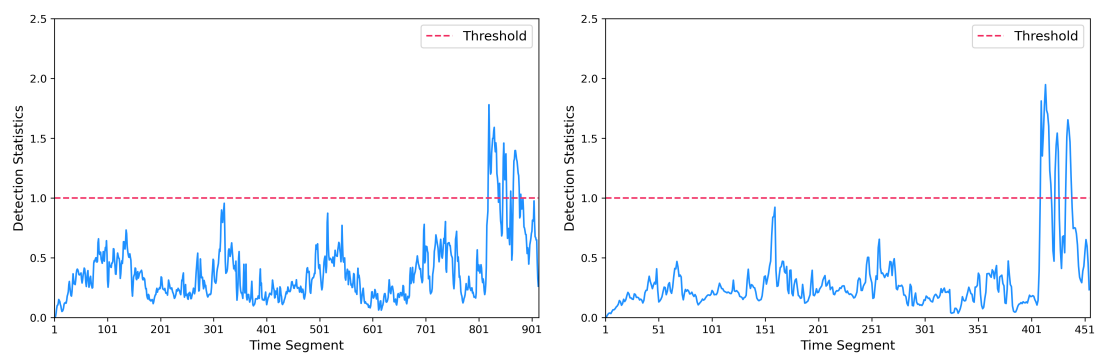

Fig 9. Degradation detection result for Penmanshiel 02

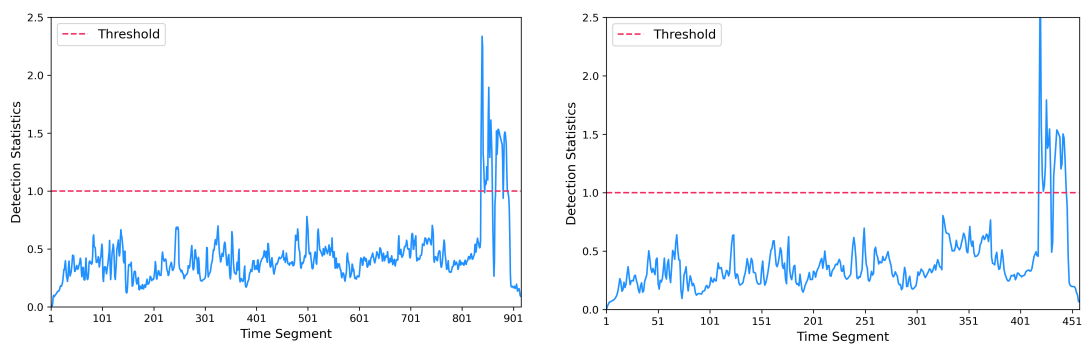

(a) $N_w = 500$          (b) $N_w = 1000$

Fig 10. Degradation detection result for Penmanshiel 08

The comparison between normal power curve and degraded power curve is shown in Fig. 11 for data segment # 20, # 266, # 843, # 913 respectively when $N_w = 500$. Penmanshiel 08 occurs performance degradation in data segment #843, while in normal operation in other three data segments. Fig. 11(a) (d) compares the difference between reference power curve and normal power curve which is constructed using complete data samples. Fig. 11(b) compares difference between reference power curve and normal power curve with incomplete data samples. Fig. 11(c) compares the difference between reference power curve and degraded power curve which is constructed using data samples in data segment #843. It can be seen that degraded power curve deviates from reference power curve significantly due to decreased power generation efficiency and performance degradation leads to the movement of the whole power curve, which validates the setting of hypothesis $H_1$. In comparison, the difference between normal power curve and reference power curve is little no matter whether data samples to construct the power curve are complete or not.

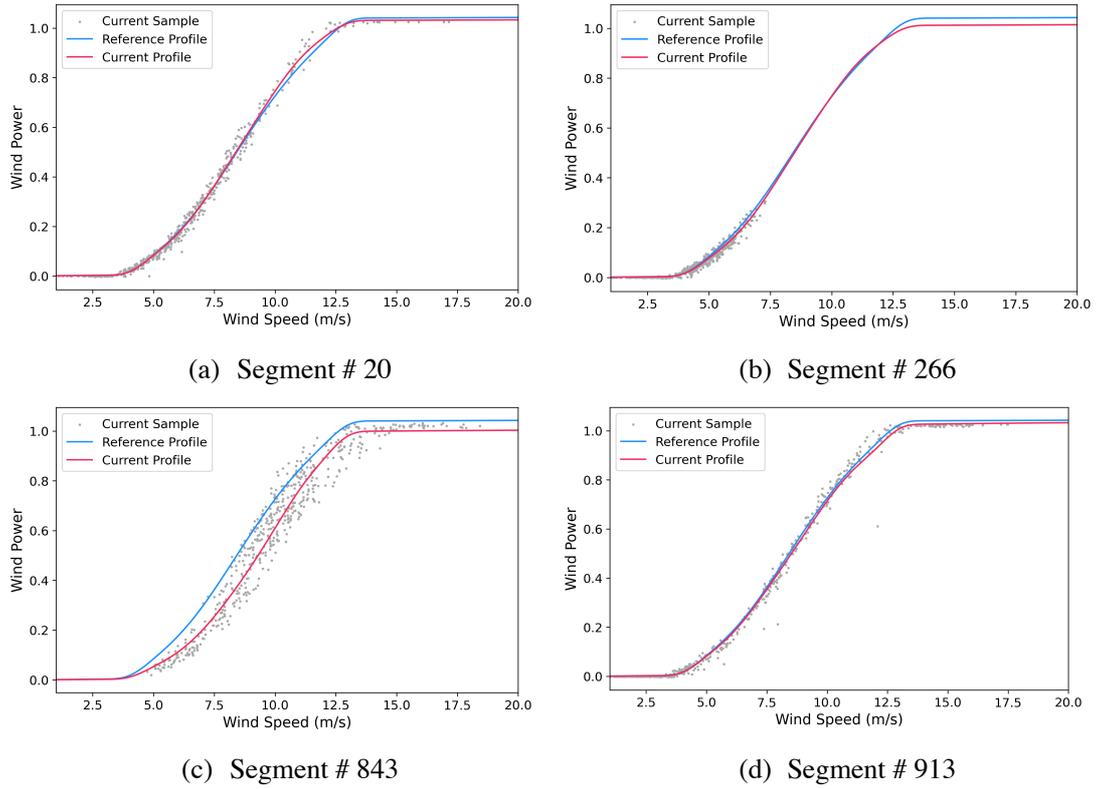

(a) Segment # 20  (b) Segment # 266

(c) Segment # 843  (d) Segment # 913

Fig 11. Power curve comparison for Penmanshiel 08 when $N_w = 500$

To further validate the performance of proposed method (denoted by CVI), we compare it with other three baselines including the chart proposed by [15] (denoted by LWZ), Hoteling T2 control chart combined with GRP (denoted by GRP) [33], and generalized likelihood ratio test combined with local linear regression (denoted by LLR) [7]. The detailed explanation of these three baselines is presented in Appendix. D. The performance degradation detection comparison of the proposed method and other three benchmarks is shown in Table 2. From Table 2, CVI outperforms LWZ, GPR, LLR in terms of precision, recall and F1 score. This validates the superiority of CVI in detecting wind turbine performance degradation. The performance degradation detection result for Penmanshiel 08 is superior to Penmanshiel 02 which is because the decline in power generation efficiency of the former wind turbine is more significant. LLR generally ranks #2 in three metrics for two WTs. Similar to WTPC modeling, the monitoring performance of GPR improves when sample size increases.

Table 2. Comparison of different degradation detection method

| $N_w$ | Method | Penmanshiel 02 | | | Penmanshiel 08 | | |
|---|---|---|---|---|---|---|---|
| | | Recall | Precision | F1 | Recall | Precision | F1 |
| 500 | CVI | **0.695** | **0.953** | **0.804** | **0.846** | 1 | **0.917** |
| | LWZ | 0.474 | 0.583 | 0.523 | 0.577 | 0.857 | 0.690 |
| | GPR | 0.559 | 0.75 | 0.641 | 0.692 | 1 | 0.818 |
| | LLR | 0.525 | 0.939 | 0.674 | 0.769 | 1 | 0.870 |
| 1000 | CVI | **0.690** | **1** | **0.816** | **0.926** | 1 | **0.962** |
| | LWZ | 0.655 | 0.760 | 0.704 | 0.852 | 1 | 0.920 |
| | GPR | 0.621 | 0.857 | 0.720 | 0.814 | 1 | 0.898 |
| | LLR | 0.620 | 0.9 | 0.734 | 0.778 | 1 | 0.875 |

# 5. Conclusion

This paper proposes a novel wind turbine performance degradation detection algorithm aimed at dealing with incomplete data sample problem. Copula-based variational inference is proposed to estimate parameters of sequential incomplete power curves and a new statistics called KL-divergence factor is proposed to detection performance degradation directionally. The prediction results of WTPC modeling and performance degradation monitoring are analyzed in two datasets. The main results are the followings:

1) Copula-based variational inference allows flexible modeling of posterior distribution and achieves stable sequential updating of model coefficient distributions given incomplete data samples.

2) WTPC modeling based on CVI and I-Spline fits data samples well and constructs reliable profiles compared with multivariate gaussian proposal because the latter underestimates distribution variance significantly. It also facilities subsequent degradation monitoring because the values of its parameters indicates WT performance.

3) The control chart generated by KL-divergence factor achieves the best monitoring performance. It can monitor wind turbines in performance degradation accurately. Moreover, it rarely triggers false alarms, which can reduce the number of unnecessary maintenances.

In the future, the proposed directional sequential incomplete power curve monitoring scheme can be expanded in the following directions. Firstly, this paper focuses on incomplete power curve monitoring task of a single WT. Multiple power curve monitoring task of different WTs is valuable to investigate. Then, profile characteristics of different components of a single WT is valuable to investigate which can facilitate fault diagnosis of WT. It can also help the understanding of WT operation.

**Acknowledgments**

This project is funded by the Natural Science Foundation of China. (72072114)

**Appendix**

A. KL divergence decomposition

Letting variational proposal in Sklar's representation be

$$q_C(\Theta) = c\left(F_1(\theta_1), \ldots, F_p(\theta_p)\right) \prod_{i=1}^{p} f_i(\theta_i)$$

and the true posterior be $p(\Theta|X) = c^*\left(F_1^*(\theta_1), \ldots, F_p^*(\theta_p)\right) \prod_{i=1}^{p} f_i^*(\theta_i)$, the KL divergence decomposes into copula terms and PDF terms,

$$KL\{q_C(\Theta)||p(\Theta|X)\} = \int q_C(\Theta)\left(\log \frac{q_C(\Theta)}{p(\Theta|X)}\right) d\Theta$$

$$= \int c(F(\Theta)) \prod_{i=1}^{p} f_i(\theta_i) \log \frac{c(F(\Theta)) \prod_{i=1}^{p} f_i(\theta_i)}{c^*(F^*(\Theta)) \prod_{i=1}^{p} f_i^*(\theta_i)} d\Theta$$

$$= \int c(F(\Theta))(\log \frac{c(F(\Theta))}{c^*(F^*(\Theta))}) \prod_{i=1}^{p} dF_i(\theta_i)$$

$$+ \int c(F(\Theta)) \prod_{i=1}^{p} f_i(\theta_i) (\log \frac{\prod_{i=1}^{p} f_i(\theta_i)}{\prod_{i=1}^{p} f_i^*(\theta_i)}) \prod_{i=1}^{p} d\theta_i$$

The first term is,

$$\int c(F(\Theta))(\log \frac{c(F(\Theta))}{c^*(F^*(\Theta))}) \prod_{i=1}^{p} dF_i(\theta_i) = KL\{c(F(\Theta))||c^*(F^*(\Theta))\}$$

The second term is,

$$\int c(F(\Theta)) \prod_{i=1}^{p} f_i(\theta_i) (\log \frac{\prod_{i=1}^{p} f_i(\theta_i)}{\prod_{i=1}^{p} f_i^*(\theta_i)}) \prod_{i=1}^{p} d\theta_i$$

$$= \sum_i \int c(F(\Theta)) \prod_{i=1}^{p} f_i(\theta_i) (\log \frac{\prod_{i=1}^{p} f_i(\theta_i)}{\prod_{i=1}^{p} f_i^*(\theta_i)}) \prod_{i=1}^{p} d\theta_i$$

$$= \sum_i \int f_i(\theta_i)(\log \frac{f_i(\theta_i)}{f_i^*(\theta_i)}) d\theta_i$$

$$= \sum_i KL\{f_i(\theta_i)||f_i^*(\theta_i)\}$$

To get line 2 to line 3 in the above equation, because $c(F(\Theta)) \prod_{i=1}^{p} f_i(\theta_i)$ is the joint distribution of $\{\theta_1, \ldots, \theta_p\}$, $\int (F(\Theta)) \prod_{i=1}^{p} f_i(\theta_i) \prod_{i \neq k} d\theta_i = f_k(\theta_k)$.

Therefore,

$$KL\{q_C(\Theta)||p(\Theta|X)\} = KL\{c(F(\Theta))||c^*(F^*(\Theta))\} + \sum_i KL\{f_i(\theta_i)||f_i^*(\theta_i)\}$$

B. Reparametrized ELBO

If $h(\cdot)$ is a non-decreasing function, then ELBO is calculated,

$$J(q) = \ln p(X) - \int q_{VC}(\Theta) \log \frac{q_{VC}(\Theta)}{p(\Theta|X)} d\Theta$$

$$= \int q_{VC}(\Theta) \log p(X) d\Theta - \int q_{VC}(\Theta) \log q_{VC}(\Theta) d\Theta + \int q_{VC}(\Theta) \log p(\Theta|X) d\Theta$$

$$= \int q_c(h^{-1}(\Theta); \Psi) \cdot \prod_{i=1}^{p} \frac{d}{d\theta_i} h_i^{-1}(\theta_i) \cdot \log p(X, \Theta) d\Theta$$

$$- \int q_c(h^{-1}(\Theta); \Psi) \cdot \prod_{i=1}^{p} \frac{d}{d\theta_i} h_i^{-1}(\theta_i) \log q_c(h^{-1}(\Theta); \Psi) \cdot \prod_{i=1}^{p} \frac{d}{d\theta_i} h_i^{-1}(\theta_i) d\Theta$$

The first term is,

$$\int q_c(h^{-1}(\Theta); \Psi) \cdot \prod_{i=1}^{p} \frac{d}{d\theta_i} h_i^{-1}(\theta_i) \cdot \log p(X, \Theta) d\Theta$$

$$= \int q_c(Z; \Psi) \cdot \prod_{i=1}^{p} \frac{1}{h_i'(z_i)} \log p(h(Z), X) \prod_{i=1}^{p} h_i'(z_i) \, dZ$$

$$= \int q_c(Z; \Psi) \cdot \log p(h(Z), X) dZ$$

The second term is,

$$\int q_c(h^{-1}(\Theta); \Psi) \cdot \prod_{i=1}^{p} \frac{d}{d\theta_i} h_i^{-1}(\theta_i) \log \left[ q_c(h^{-1}(\Theta); \Psi) \cdot \prod_{i=1}^{p} \frac{d}{d\theta_i} h_i^{-1}(\theta_i) \right] d\Theta$$

$$= \int q_c(h^{-1}(\Theta); \Psi) \prod_{i=1}^{p} \frac{d}{d\theta_i} h_i^{-1}(\theta_i) \log q_c(h^{-1}(\Theta); \Psi) d\Theta$$

$$+ \int q_c(h^{-1}(\Theta); \Psi) \prod_{i=1}^{p} \frac{d}{d\theta_i} h_i^{-1}(\theta_i) \log \prod_{i=1}^{p} \frac{d}{d\theta_i} h_i^{-1}(\theta_i)] d\Theta$$

$$= \int q_c(Z; \Psi) \log q_c(Z; \Psi) dZ - \int q_c(Z; \Psi) \log h'(Z) dZ$$

Therefore,

$$J(q) = \mathbb{E}_{q_c(Z;\Psi)} \left[ \log p(h(Z), X) - \log q_c(Z; \Psi) + \sum_{i=1}^{p} \log h_i'(z_i) \right]$$

Similarly, ff $h(\cdot)$ is a monotonous decreasing function, then ELBO is represented,

$$J(q) = (-1)^p \mathbb{E}_{q_c(Z;\Psi)} \left[ \log p(h(Z), X) - \log q_c(Z; \Psi) + \sum_{i=1}^{p} \log |h_i'(z_i)| \right]$$

C. MGR framework

Multivariate Gaussian proposal (MGR) uses Gaussian copula to model dependence structure of latent variables and Gaussian distribution as marginal distribution. Giving the training samples of size N, the I-spline regression model is written as the matrix form,

$$Y = Z\beta + e$$

where $Y, e \in \mathcal{R}^N$; $\beta \in \mathcal{R}^{K+p+1}$ and $Z \in \mathcal{R}^{N \times (K+p+1)}$.

In MGR framework, the profile construction and updating algorithm is summarized as follows. The prior distribution on regression coefficient $\beta$ is assumed to be multivariate Gaussian distribution with mean $\mu_0$ and covariance matrix $\Sigma_0$,

$$p(\beta) = \mathcal{N}(\mu_0, \Sigma_0)$$

Where $\mathcal{N}(\cdot)$ denotes the Gaussian distribution. The error term $e_i$ is assumed to obey a Gaussian distribution with mean 0 and variance $\sigma_i^{-1}$, which is given a Gamma prior distribution to complete the Bayesian model,

$$p(e_i|\sigma_i) = \mathcal{N}(e_i|0, \sigma_i^{-1}), \quad p(\sigma_i) = \mathcal{G}(\sigma_i|a_0, b_0)$$

The corresponding parameters can be represented as $\Theta = \{\beta, \sigma\}$, where $\beta = \{\beta_1, \beta_2, \ldots, \beta_{K+p+1}\}$, $\sigma = \{\sigma_1, \sigma_2, \ldots, \sigma_N\}$. The posterior distribution of $\Theta$ can be represented as:

$$P(\Theta) = P(\beta) \prod_{i=1}^{N} p(\sigma_i)$$

$q(\beta)$ is assumed to be multivariate Gaussian distribution, $q(\beta) = \mathcal{N}(\beta|\mu_\beta, \Sigma_\beta)$; $q(\sigma_i)$ is assumed to be Gamma distribution, $q(\sigma_i) = \mathcal{G}(\sigma_i|a_i, b_i)$. After representing $E_{q(\Theta)} lnp(Y, \Theta)$ and $E_{q(\Theta)} lnq(\Theta)$, ELBO $J(\tilde{p})$ can be represented by,

$$J(\tilde{p}) \propto \frac{1}{2} \sum_{i=1}^{N} <ln\sigma_i> - \frac{1}{2} \sum_{i=1}^{N} <\sigma_i> \left(y_i^2 - 2y_i Z_i^T \mu_\beta + \mu_\beta^T Z_i Z_i^T \mu_\beta + tr(Z_i Z_i^T \Sigma_\beta)\right)$$

$$-\frac{M}{2} ln 2\pi - \frac{1}{2} ln|\Sigma_0| - \frac{1}{2}\left(\mu_\beta^T \Sigma_0^{-1} \mu_\beta + tr(\Sigma_0^{-1} \Sigma_\beta) - 2\mu_0^T \Sigma_0^{-1} \mu_\beta + \mu_0^T \Sigma_0^{-1} \mu_0\right)$$

$$+(a_0 - 1) \sum_{i=1}^{N} <ln\sigma_i> - b_0 \sum_{i=1}^{N} <\sigma_i> + \frac{1}{2} ln|\Sigma_\beta| - \sum_{i=1}^{N} a_i lnb_i$$

$$-\sum_{i=1}^{N}(a_i - 1)<ln\sigma_i> + \sum_{i=1}^{N} ln\Gamma(a_i) + \sum_{i=1}^{N} b_i <\sigma_i>$$

where $<ln\sigma_i> = \psi(a_i) - lnb_i$; $\psi(x)$ is the digamma function, $\psi(x) = \frac{dln\Gamma(x)}{dx}$; $<\sigma_i> = a_i/b_i$.

To solve the optimization problem,

$$\frac{\partial J(\tilde{p})}{\partial \Sigma_\beta} = -\frac{1}{2} \sum_{i=1}^{N} <\sigma_i> Z_i Z_i^T - \frac{1}{2}(\Sigma_0^{-1})^T + \frac{1}{2}\frac{1}{|\Sigma_\beta|}|\Sigma_\beta|(\Sigma_\beta^{-1})^T = 0$$

$$\frac{\partial J(\tilde{p})}{\partial \boldsymbol{\mu}_\beta} = \boldsymbol{Z}^T diag<\boldsymbol{\sigma}>\boldsymbol{Y} - \boldsymbol{Z}^T diag<\boldsymbol{\sigma}>\boldsymbol{Z}\boldsymbol{\mu}_\beta - \boldsymbol{\Sigma}_0^{-1}\boldsymbol{\mu}_\beta + \boldsymbol{\Sigma}_0^{-1}\boldsymbol{\mu}_0 = 0$$

$$\frac{\partial J(\tilde{p})}{\partial a_i} = \frac{d\psi(a_i)}{da_i}\left(\frac{1}{2} - a_i + a_0\right) - \frac{b_0 + \frac{1}{2}<(y_i - Z_i\boldsymbol{\beta})^2>}{b_i} + 1 = 0$$

$$\frac{\partial J(\tilde{p})}{\partial b_i} = -\frac{a_0 + \frac{1}{2}}{b_i} + \frac{a_i(b_0 + <(y_i - Z_i\boldsymbol{\beta})^2>)}{b_i^2} = 0$$

Solving the optimization problem, we can obtain,

$$\boldsymbol{\Sigma}_\beta = [\boldsymbol{\Sigma}_0^{-1} + \boldsymbol{Z}^T diag(<\boldsymbol{\sigma}>)\boldsymbol{Z}]^{-1}$$

$$\boldsymbol{\mu}_\beta = \boldsymbol{\Sigma}_\beta[\boldsymbol{Z}^T diag(<\boldsymbol{\sigma}>)\boldsymbol{Y} + \boldsymbol{\Sigma}_0^{-1}\boldsymbol{\mu}_0]$$

$$a_i = a_0 + \frac{1}{2}, b_i = b_0 + \frac{1}{2}<(y_i - Z_i\boldsymbol{\beta})^2>$$

Because $\boldsymbol{Z}^T diag(<\boldsymbol{\sigma}>)\boldsymbol{Z}$ is always positive, $\boldsymbol{\Sigma}_\beta^{-1}$ is always bigger than $\boldsymbol{\Sigma}_0^{-1}$, which indicates why the variance of posterior distribution inferenced by MGR framework is severely underestimated during sequential updating.

D. Three performance degradation detection baselines
(1) LWZ

Firstly, the profile function is modeled by WCDF, that is $P = 1 - e^{-(x/c)^k}$ when $v_c < v < v_r$; $P = 0$ when $v < v_c$; $P = 1$ when $v > v_c$. Then, it is obvious that the two parameters in WCDF are dependent. To monitor the two parameters simultaneously, Hotelling's T2 control chart is applied. The detailed explanation can be seen in paper [15].

(2) GPR

A GP is a collection of random variables with a joint Gaussian distribution. We have $p_i$ observations from sample $n$ of profile $i$, we can consider them as a sample of a $p_i$-variate Gaussian distribution. A GP model for profile $i$ is completely specified by its mean function $\mu_i(x_{i,j})$ and covariance function $\mathcal{K}_i(x_{i,j}, x_{i,j'})$, the GP regression model for profile $i$ can be expressed as follows:

$$y_i(x_{i,j}) = f_i(x_{i,j}) + \varepsilon_i(x_{i,j})$$

$$f_i(x_{i,j}) \sim GP\left(\mu_i(x_{i,j}), \mathcal{K}_i(x_{i,j}, x_{i,j'})\right), \quad \varepsilon_i \sim N(0, \sigma^2)$$

where $\varepsilon_i$ represents the measurement noise; $\mu_i(x_{i,j})$ is usually taken to be zero; the model is parameterized by defining a positive definite covariance function, which is chosen to be a Gaussian kernel. The prediction for each design point $j$ of profile $i$,

$$\hat{y}_i(x_{i,j})|\boldsymbol{y}_i = \mathcal{N}(\boldsymbol{\eta}_i^T(x_{i,j})(\boldsymbol{\Sigma}_i)^{-1}\boldsymbol{y}_i, \Sigma_i(x_{i,j}, x_{i,j}) - \boldsymbol{\eta}_i^T(x_{i,j})(\boldsymbol{\Sigma}_i)^{-1}\boldsymbol{\eta}_i(x_{i,j}))$$

where $\boldsymbol{\eta}_i(x_{i,j}) = [\Sigma_i(x_{i,j}, x_{i,1}), \Sigma_i(x_{i,j}, x_{i,2}), \ldots, \Sigma_i(x_{i,j}, x_{i,p_i})]^T$, $\Sigma_i(x_{i,j}, x_{i,j}) = \mathcal{K}_i(x_{i,j}, x_{i,j}) + \sigma^2$. Using this method, we can fit a GP model for each profile $i$. Then, we need to estimate the IC profile $i$ by pooling all fitted GP models. In this regard, for each profile $i$, the final vector of predicted values in its design points $\hat{\boldsymbol{y}}_i^*$ and $\boldsymbol{\Sigma}_i^*$ can be estimated.

Having estimated the required process parameters of each profile $i$ using the GP regression model, we can establish univariate control chart for profile $i$,

$$T_i^2 = (\mathbf{y}_i - \hat{\mathbf{y}}_i^*)^T (\mathbf{\Sigma}_i^*)^{-1} (\mathbf{y}_i - \hat{\mathbf{y}}_i^*)$$

This control chart triggers an OC signal if $T_i^2 > L_i$, where $L_i > 0$ is the control limit chosen to satisfy,

$$\Pr(T_i^2 > L_i) = \alpha_i$$

when the process is IC. The value $\alpha_i$ is called the false alarm rate.

(3) LLR

LLR method is based on nonparametric regression and generalized likelihood ratio test. The underlying model for profile is,

$$y_{ij} = g(x_{ij}) + \varepsilon_{ij} \quad i = 1, \ldots, n, j = 1, 2, \ldots$$

where $x_{ij}$ denotes the value of the regressor for the $i$th observation in the $j$th profile; $g(\cdot)$ denotes local linear smoother function. LLR method consider the Phase II case in which the IC regression function and variance, say $g_0$ and $\sigma_0^2$, are assumed to be known; that is, it is assumed that the IC dataset used in Phase I is sufficient to estimate the regression model well.

To monitor a general profile model (2), the regression function, $g$, and the standard deviations, $\sigma$, should be controlled simultaneously. The nonparametric monitoring approach for $g$ is based on the Generalized Likelihood Ratio (GLR) statistics proposed by paper [34]. The null hypothesis testing problem is,

$$H_0: g = g_0, \quad \sigma = \sigma_0$$
$$H_1: g \neq g_0, \quad \sigma = \sigma_0$$

The GLR test statistic is,

$$lr = \frac{1}{\sigma_0^2} [\sum_{i=1}^{n} (y_i - g_0(x_i))^2 - (y_i - \hat{g}(x_i))^2]$$

where $\hat{g}$ is a reasonable nonparametric estimator that leads to the logarithm of likelihood function $H_1$ to replace the unknown function $g$ under $H_1$.